\documentclass[submission,copyright,creativecommons]{eptcs}
\providecommand{\event}{TERMGRAPH 2020} 

\usepackage{graphicx}
\graphicspath{{pics/}}
\usepackage{amsmath}
\usepackage{amsthm}
\usepackage{wrapfig}

\usepackage{listings}
\usepackage{xcolor}

\definecolor{codegreen}{rgb}{0,0.6,0}
\definecolor{codegray}{rgb}{0.5,0.5,0.5}	
\definecolor{codepurple}{rgb}{0.58,0,0.82}
\definecolor{backcolour}{rgb}{0.95,0.95,0.92}

\lstdefinestyle{mystyle}{
	basicstyle=\ttfamily\footnotesize,
	breakatwhitespace=false,         
	breaklines=true,                 
	captionpos=b,                    
	keepspaces=true,                 
	numbers=left,                    
	numbersep=5pt,                  
	showspaces=false,                
	showstringspaces=false,
	showtabs=false,                  
	tabsize=2
}
\title{Operational Semantics with\\Hierarchical Abstract Syntax Graphs\footnote{Extended abstract of invited talk}}
\author{Dan R. Ghica
\institute{Huawei Research, Edinburgh}
\institute{University of Birmingham, UK}
}
\def\titlerunning{Operational Semantics with Hierarchical ASGs}
\def\authorrunning{Dan R. Ghica}
\begin{document}
\maketitle

\begin{abstract}
This is a motivating tutorial introduction to a semantic analysis of programming languages using a graphical language as the representation of terms, and graph rewriting as a representation of reduction rules. 
We show how the graphical language automatically incorporates desirable features, such as $\alpha$-equivalence and how it can describe pure computation, imperative store, and control features in a uniform framework. 
The graph semantics combines some of the best features of structural operational semantics and abstract machines, while offering powerful new methods for reasoning about contextual equivalence. 

All technical details are available in an extended technical report by Muroya and the author~\cite{DBLP:journals/corr/abs-1907-01257} and in Muroya's doctoral dissertation~\cite{koko}. 
\end{abstract}

\section{Hierarchical abstract syntax graphs}

Before proceeding with the business of analysing and transforming the source code of a program, a compiler first parses the input text into a sequence of atoms, the \textit{lexemes}, and then assembles them into a tree, the \textit{Abstract Syntax Tree} (AST), which corresponds to its grammatical structure. 
The reason for preferring the AST to raw text or a sequence of lexemes is quite obvious.
The structure of the AST  incorporates most of the information needed for the following stage of compilation, in particular identifying operations as nodes in the tree and operands as their branches. 
This makes the AST algorithmically well suited for its purpose.
Conversely, the AST excludes irrelevant lexemes, such as separators (white-space, commas, semicolons) and aggregators (brackets, braces), by making them implicit in the tree-like structure. 
It is always possible to recover the textual input, or rather an equivalent version of it, from the AST via a process known as \textit{pretty-printing}. 

A fair question to ask is whether the AST can be improved upon as a representation of program text, which captures grammatical structure while discarding needless detail.
In pretty-printing we know how irrelevant lexemes can be manipulated to achieve a certain aesthetic effect. 
Redundant brackets can be elided to reduce clutter, white-space can be added or removed to improved alignment, and so on. 
Such details are accepted as irrelevant.

There is another, deeper level of detail in the text, which is irrelevant but not always appreciated as such: variable names. 
Whereas we know, formally, that bound variables can be systematically renamed (\textit{$\alpha$-equivalence}) the conventional AST will still remember their names, even though variable names induce bureaucratic complications having to do with scope and shadowing. 
Finally, there is yet another even deeper level of irrelevant detail in the program text, the order in which variables are defined, absent define-use dependencies between the definitions. 

\begin{figure}
\centering 
{\small
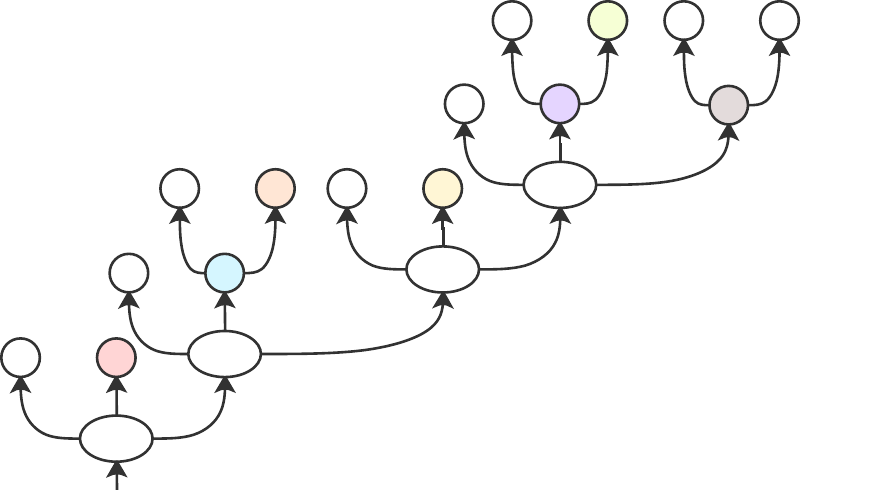
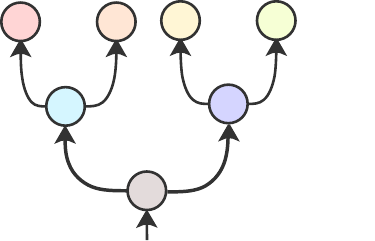
}
\caption{AST vs ASG for variable definition}
\label{fig:ast}
\end{figure}

Consider the program text in Fig.~\ref{fig:cwb}, in which \texttt{def} is a binder associating a variable with a definition as text, akin to a macro, but respecting variable scoping rules. 
Variable \textit{x} in line 3 could be renamed, on lines 3-5, to something else to avoid shadowing the variable with the same name on line 1. 
Lines 1-2 and lines 3-4 can be swapped without changing the result. 
But these facts are not immediate from examining its AST in Fig~\ref{fig:ast} (left). 

\begin{wrapfigure}{l}{0.35\textwidth}
	\begin{lstlisting}[language=Python]
	def x = 0
	def y = x + 1
	def x = 2
	def z = x + 3
	y + z\end{lstlisting}
	\caption{Bindings}
	\label{fig:cwb}
\end{wrapfigure}
Unlike ASTs, an abstract syntax graphs (ASG) do not treat binders and variables as nodes. 
Variables are instead represented as links, and binders assign target nodes corresponding to their definitions to the variable links. 
The ASG of the code in Fig.~\ref{fig:cwb} is represented next to its AST in Fig.~\ref{fig:ast}.
To better understand the relation between the AST and the ASG, corresponding nodes are coloured and the links are labelled with variable names.
The colour and the labels are not part of the definition of the graph structure, but are just used to aid understanding. 
The nodes corresponding to variable uses and definitions, left blank, are not part of the ASG. 
It is thus immediately obvious that the ASG is, by construction, quotiented both by $\alpha$-equivalence and by the order of non-interfering variable bindings.
This more general structural equivalence of lambda calculus terms has been dubbed ``graph equivalence" by Accattoli et. al.~\cite{DBLP:conf/popl/AccattoliBKL14}.

\begin{wrapfigure}{l}{0.35\textwidth}
	\begin{lstlisting}[language=Python]
	def x = 2
	def z = x + 3
	z + z\end{lstlisting}
	\caption{Contraction}
	\label{fig:con}
\end{wrapfigure}
ASGs are also helpful when variables are reused, as in the Fig.~\ref{fig:con} example. 
The AST and the ASG are showed side-by-side in Fig.~\ref{fig:ast}, noting that the links corresponding to the variable \texttt{z}, used twice, now both point to its definition. 
This is why ASG are no longer trees, but directed acyclic graphs (DAGs). 
Formally, the ASGs are \textit{hypergraphs} with the links a graphical representation of vertices and the nodes a graphical representation of hyperedges.
\begin{figure}
	\centering
	{\small
	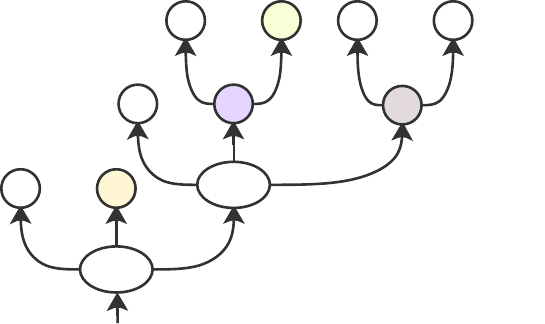
\begingroup%
  \makeatletter%
  \providecommand\color[2][]{%
    \errmessage{(Inkscape) Color is used for the text in Inkscape, but the package 'color.sty' is not loaded}%
    \renewcommand\color[2][]{}%
  }%
  \providecommand\transparent[1]{%
    \errmessage{(Inkscape) Transparency is used (non-zero) for the text in Inkscape, but the package 'transparent.sty' is not loaded}%
    \renewcommand\transparent[1]{}%
  }%
  \providecommand\rotatebox[2]{#2}%
  \newcommand*\fsize{\dimexpr\f@size pt\relax}%
  \newcommand*\lineheight[1]{\fontsize{\fsize}{#1\fsize}\selectfont}%
  \ifx\svgwidth\undefined%
    \setlength{\unitlength}{61.38549042bp}%
    \ifx\svgscale\undefined%
      \relax%
    \else%
      \setlength{\unitlength}{\unitlength * \real{\svgscale}}%
    \fi%
  \else%
    \setlength{\unitlength}{\svgwidth}%
  \fi%
  \global\let\svgwidth\undefined%
  \global\let\svgscale\undefined%
  \makeatother%
  \begin{picture}(1,1.33078956)%
    \lineheight{1}%
    \setlength\tabcolsep{0pt}%
    \put(0,0){\includegraphics[width=\unitlength,page=1]{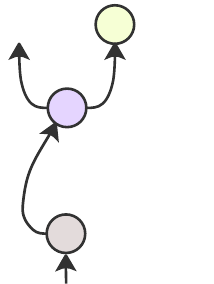}}%
    \put(0.27117499,0.78702782){\makebox(0,0)[lt]{\lineheight{1.25}\smash{\begin{tabular}[t]{l}\texttt{+}\end{tabular}}}}%
    \put(0.27440806,0.19603557){\makebox(0,0)[lt]{\lineheight{1.25}\smash{\begin{tabular}[t]{l}\texttt{+}\end{tabular}}}}%
    \put(0.507133,1.18551369){\makebox(0,0)[lt]{\lineheight{1.25}\smash{\begin{tabular}[t]{l}\texttt{3}\end{tabular}}}}%
    \put(0,0){\includegraphics[width=\unitlength,page=2]{asgcon.pdf}}%
    \put(0.05965534,1.20011675){\makebox(0,0)[lt]{\lineheight{1.25}\smash{\begin{tabular}[t]{l}\texttt{2}\end{tabular}}}}%
    \put(0,0){\includegraphics[width=\unitlength,page=3]{asgcon.pdf}}%
    \put(-0.00739754,0.39199088){\makebox(0,0)[lt]{\lineheight{1.25}\smash{\begin{tabular}[t]{l}$z$\end{tabular}}}}%
    \put(0.55681321,0.38606386){\makebox(0,0)[lt]{\lineheight{1.25}\smash{\begin{tabular}[t]{l}$z$\end{tabular}}}}%
    \put(0.13204197,0.93925441){\makebox(0,0)[lt]{\lineheight{1.25}\smash{\begin{tabular}[t]{l}$x$\end{tabular}}}}%
  \end{picture}%
\endgroup%

	}
	\caption{AST vs ASG for contraction}
	\label{fig:asg}
\end{figure}

To represent \textit{local} variable binding, as encountered in functions as opposed to simple variable \textit{definition} discussed above, we note that local variable binding is always associated with \textit{thunks}, i.e. code with delayed execution.
This is related to the fact that conventional programming languages use normal-order reduction. 
In this evaluation strategy functions are considered values, i.e. there is no evaluation `under the lambda'.
In other words, functions are thunks and locally-bound variables will always induce a thunk. 
Because thunks can be considered, operationally, as a single entity, it is convenient to represent them in their ASG form as a single node, labeled by the definition of the thunk, which is also an ASG. 
In other words, to model local variable binding in functions it is convenient to use graphs labelled by graphs, which are, formally, \textit{hierarchical} hypergraphs. 

\begin{figure}
	\centering\small 
	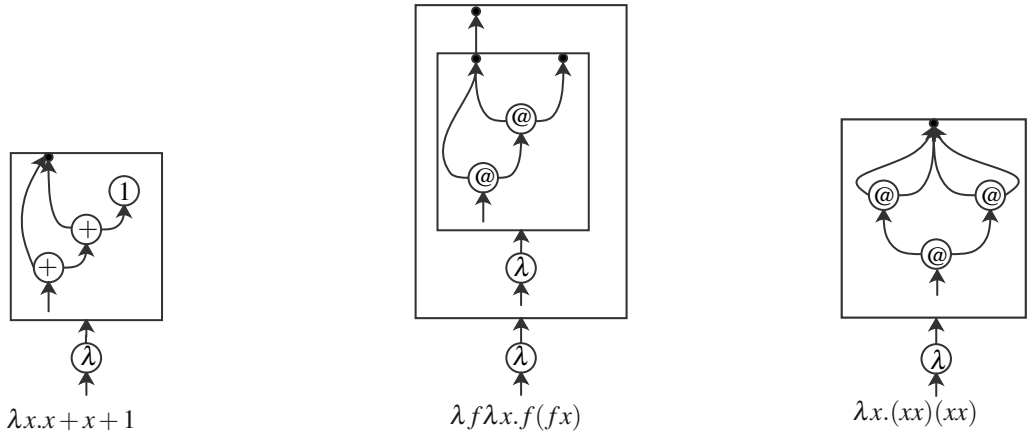
	\caption{Hierarchical ASGs}
	\label{fig:hasg}
\end{figure}

To model variable behaviour in thunks correctly, our ASGs need to be \textit{interfaced}, i.e. there needs to be a defined order on incoming links. 
If a thunk has $m$ bound variables and $n$ free variables then the first $m$ incoming links of the ASG used as its label represent the bound variables, in the order in which they are bound. 
The last $n$ incoming links represent the free variables, in some specified order. 
The node corresponding to the thunk will also have $n$ incoming links, representing the definitions of its $n$ free variables, in an order consistent with the order used by the label ASG. 
To make the correspondence more perspicuous we connect the links corresponding to the free variables from the labelling ASG to those of the node, as it causes no ambiguity. 
Fig.~\ref{fig:hasg} shows several examples for hierarchical ASGs and the corresponding terms, with function application labelled as $@$.
Note that thunks associated with lambda expressions are still explicitly linked to a lambda-labelled node. 
This is because in a programming language thunks can be used for expressions other than function definitions, as we shall see. 

\section{Operational semantics}

The most widely used method for specifying programming languages is via \textit{operational semantics} (OS). 
There are several versions of OS. 
We will focus on so-called \textit{structural} operational semantics (SOS) in the style of Plotkin~\cite{DBLP:journals/jlp/Plotkin04a}, in which a \textit{transition relation} is defined on \textit{configurations} consisting of a term $t$ and some additional information (e.g. a program \textit{store}, $s, t$), so that the definition of the relation is inductive on the structure of the term.

\begin{wrapfigure}{r}{.35\textwidth}
	\begin{gather*} 
	s,1+2\to s,3 \\
	\frac{s,e_1\to s',e_1'}{s,e_1+e_2\to s', e_1'+e_2}. 
	\end{gather*}
	\caption{SOS (example rules)}
	\label{fig:ltr}
\end{wrapfigure}
Typically the transition relation is written as $s,t\to s', t'$. 
There are two kinds of rules, \textit{basic reductions} which perform operations (e.g. first rule in Fig.~\ref{fig:ltr}) and \textit{simplification steps} which seek redexes structurally in the program text according to the evaluation strategy (second rule in Fig.~\ref{fig:ltr}). 
The latter are usually written in natural-deduction style. 
For example, the rule specifying that $+$ is evaluated left-to-right is the second rule in Fig.~\ref{fig:ltr}.
Note how the first operand $e_1$ is evaluated to $e_1'$ and, in the proces, the store $s$ may change to $s'$.  

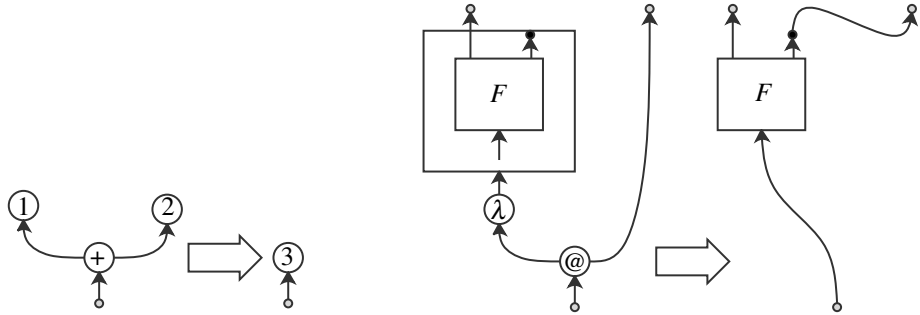
\begin{figure}
	\centering
	\small
\begingroup%
  \makeatletter%
  \providecommand\color[2][]{%
    \errmessage{(Inkscape) Color is used for the text in Inkscape, but the package 'color.sty' is not loaded}%
    \renewcommand\color[2][]{}%
  }%
  \providecommand\transparent[1]{%
    \errmessage{(Inkscape) Transparency is used (non-zero) for the text in Inkscape, but the package 'transparent.sty' is not loaded}%
    \renewcommand\transparent[1]{}%
  }%
  \providecommand\rotatebox[2]{#2}%
  \newcommand*\fsize{\dimexpr\f@size pt\relax}%
  \newcommand*\lineheight[1]{\fontsize{\fsize}{#1\fsize}\selectfont}%
  \ifx\svgwidth\undefined%
    \setlength{\unitlength}{343.10178285bp}%
    \ifx\svgscale\undefined%
      \relax%
    \else%
      \setlength{\unitlength}{\unitlength * \real{\svgscale}}%
    \fi%
  \else%
    \setlength{\unitlength}{\svgwidth}%
  \fi%
  \global\let\svgwidth\undefined%
  \global\let\svgscale\undefined%
  \makeatother%
  \begin{picture}(1,0.33882074)%
    \lineheight{1}%
    \setlength\tabcolsep{0pt}%
    \put(0,0){\includegraphics[width=\unitlength,page=1]{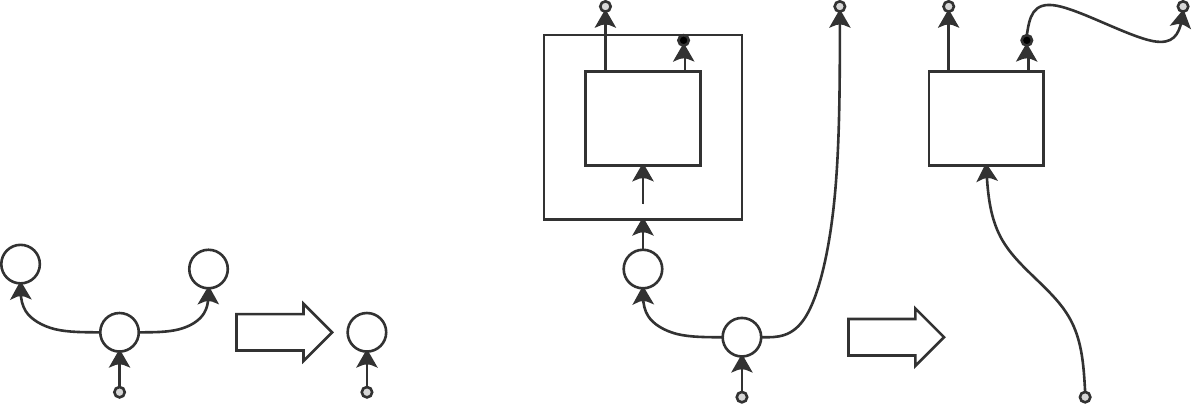}}%
    \put(0.01028763,0.10652072){\makebox(0,0)[lt]{\lineheight{1.25}\smash{\begin{tabular}[t]{l}1\end{tabular}}}}%
    \put(0.16883817,0.1066565){\makebox(0,0)[lt]{\lineheight{1.25}\smash{\begin{tabular}[t]{l}2\end{tabular}}}}%
    \put(0.09063742,0.05219257){\makebox(0,0)[lt]{\lineheight{1.25}\smash{\begin{tabular}[t]{l}+\end{tabular}}}}%
    \put(0.30002097,0.05219257){\makebox(0,0)[lt]{\lineheight{1.25}\smash{\begin{tabular}[t]{l}3\end{tabular}}}}%
    \put(0.52978201,0.23131839){\makebox(0,0)[lt]{\lineheight{1.25}\smash{\begin{tabular}[t]{l}$F$\end{tabular}}}}%
    \put(0.82086146,0.2323929){\makebox(0,0)[lt]{\lineheight{1.25}\smash{\begin{tabular}[t]{l}$F$\end{tabular}}}}%
    \put(0.53011556,0.10288978){\makebox(0,0)[lt]{\lineheight{1.25}\smash{\begin{tabular}[t]{l}$\lambda$\end{tabular}}}}%
    \put(0.60925507,0.04479491){\makebox(0,0)[lt]{\lineheight{1.25}\smash{\begin{tabular}[t]{l}$@$\end{tabular}}}}%
  \end{picture}%
\endgroup%

	\caption{ASG rewrite as basic reductions}
	\label{fig:ltrg}
\end{figure}

SOS can be naturally formulated on ASGs rather than on terms. 
Basic reductions correspond to graph rewrites and simplification steps to a graph traversal algorithm which seeks the redexes. 
The basic reduction in Fig.~\ref{fig:ltr} is shown as a graph rewrite in Fig.~\ref{fig:ltrg}, along with the rule for $\beta$-reduction. 
The former is quite obvious, but the latter is more interesting.
It consists of the deletion of the abstraction-application pair, the `unboxing' of the thunk by extracting the label of the thunk node and using it in the top-level graph, the re-wiring of the bound variable, now open, to the argument, and using the root node of $F$ as the overall root node. 
For the $\beta$ rule the graphs involved in the rewrite must also be interfaced, with the interface nodes highlighted in grey. 
Also note that the rule is actually the \textit{small} $\beta$ rule used in calculi of explicit substitution which reduces $(\lambda x.F)M$ to $\mathtt{def}\ x=M\ \mathtt{in}\ F$~\cite{DBLP:journals/jfp/AbadiCCL91}.

One aspect of the ASG-based evaluation which needs to be clearly explicated is sharing.
The DAG structure of the ASG can be refined by introducing special sharing nodes, which, unlike operation nodes, would be allowed to have multiple incoming links. 
Sharing nodes have a special behaviour during evaluation, managing the process of systematic copying of sub-graphs. 

To evaluate an ASG in a way that is consistent with left-to-right call-by-value the traversal is depth-first and left-to-right, without reaching inside thunks, starting from the unique root. 
The current link in the traversal is called the \textit{focus}, and it can move \textit{up} (i.e. away from the root) or \textit{down} (i.e. towards the root).
When the focus is moving up and it encounters a copy node it will copy the node shared by the copy node, inserting further copy nodes on its outgoing links. 
As the focus is moving down, whenever it passes a node which has an associated rewrite rule it will exercise it, then change direction and move up again (\textit{refocussing}). 
\begin{figure}[t]
	\centering\small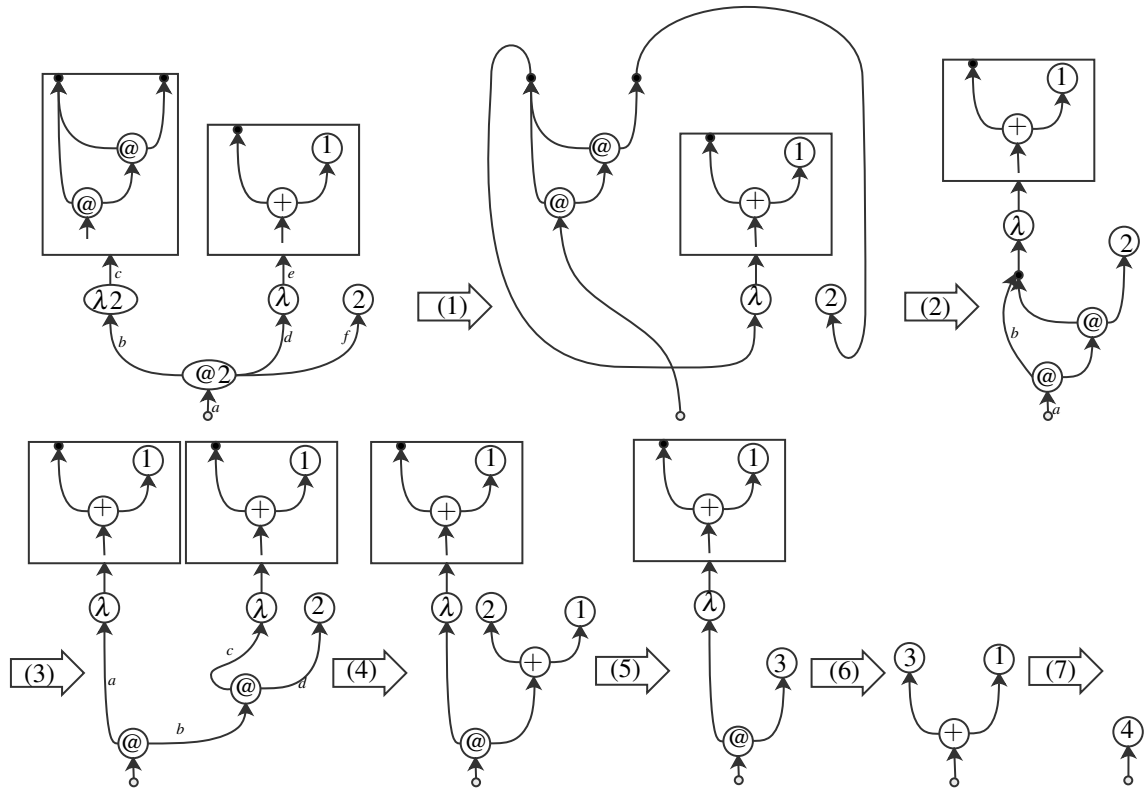
	\caption{Evaluation of $(\lambda f x.f(fx)))(\lambda x.x+1, 2)$.}
	\label{fig:exrew}
\end{figure}

In Fig.~\ref{fig:exrew} we show the key steps in the evaluation of the expression $(\lambda f x.f(fx)))(\lambda x.x+1, 2)$.
We use the labels of $\lambda2$ and $@2$ for definition and use of a function with two arguments. 
Step (1) is reached after the focus moves along the path $\mathit{abcbdefa}$, at which point the rewrite is performed, unboxing the thunk and attaching the arguments to the nodes made available. 
Step (2) is simply rearranging the ASG in a more readable format. 
Step (3) is the copying of the node corresponding to the function $\lambda x.x+1$ after the focus traverses path $ab$. 
Step (4) is the $\beta$ rewrite applied after the focus traverses path $abcdb$. 
Step (5) is an arithmetic reduction, followed by another $\beta$ rewrite and a final arithmetic reduction.

The examples in this section (abstraction, application, arithmetic) deal with what is usually deemed \textit{pure} functional programming, case in which the configuration used by the SOS is the term itself. 
Expanding the SOS of a language to incorporate effects usually requires revising the format of the configuration of the SOS, which in turn requires reformulating the rules for the preexisting operations. 
This is a major fragility of the SOS approach, since the revision of the format invalidates any technical results and require laborious re-proving~\cite{DBLP:journals/entcs/GhicaT12}.
ASGs can be enhanced with a single new node which will allow the formulation of most known effects, namely an \textit{atom} node, in the sense of~\cite{pitts_2013}.
The ASG OS for a pure language then only differs from the ASG OS of an impure language in that the atom nodes are not involved.
The atom node, just like a sharing node, allows multiple incoming links.
However, during evaluation, the atom node does not trigger a copying of the node at the end of its outgoing link, but is instead treated as an endpoint by the ASG traversal strategy.
Indeed, just as computations are not performed inside of thunks they are also not performed inside of the store.  
This insight, that the essence of effectful computation is the presence of atoms in the OS is originally due to Pitts, but it turns out to be most effective in  ASG-based OS \cite{DBLP:conf/lics/Pitts96}.  

\begin{wrapfigure}{r}{.5\textwidth}
	\centering\small 
\begingroup%
  \makeatletter%
  \providecommand\color[2][]{%
    \errmessage{(Inkscape) Color is used for the text in Inkscape, but the package 'color.sty' is not loaded}%
    \renewcommand\color[2][]{}%
  }%
  \providecommand\transparent[1]{%
    \errmessage{(Inkscape) Transparency is used (non-zero) for the text in Inkscape, but the package 'transparent.sty' is not loaded}%
    \renewcommand\transparent[1]{}%
  }%
  \providecommand\rotatebox[2]{#2}%
  \newcommand*\fsize{\dimexpr\f@size pt\relax}%
  \newcommand*\lineheight[1]{\fontsize{\fsize}{#1\fsize}\selectfont}%
  \ifx\svgwidth\undefined%
    \setlength{\unitlength}{208.5bp}%
    \ifx\svgscale\undefined%
      \relax%
    \else%
      \setlength{\unitlength}{\unitlength * \real{\svgscale}}%
    \fi%
  \else%
    \setlength{\unitlength}{\svgwidth}%
  \fi%
  \global\let\svgwidth\undefined%
  \global\let\svgscale\undefined%
  \makeatother%
  \begin{picture}(1,0.42446043)%
    \lineheight{1}%
    \setlength\tabcolsep{0pt}%
    \put(0,0){\includegraphics[width=\unitlength,page=1]{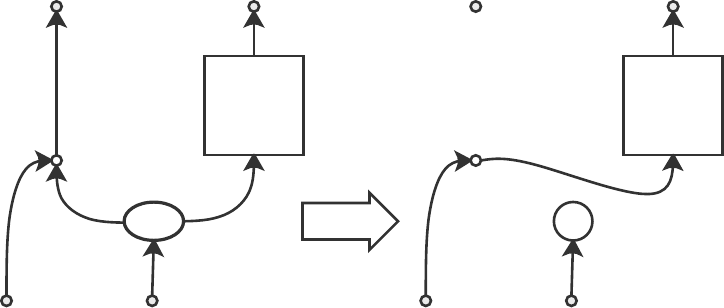}}%
    \put(0.18192153,0.10796573){\makebox(0,0)[lt]{\lineheight{1.25}\smash{\begin{tabular}[t]{l}${:}{=}$\end{tabular}}}}%
    \put(0.32448734,0.25679721){\makebox(0,0)[lt]{\lineheight{1.25}\smash{\begin{tabular}[t]{l}$M$\end{tabular}}}}%
    \put(0.78121339,0.10796573){\makebox(0,0)[lt]{\lineheight{1.25}\smash{\begin{tabular}[t]{l}$\bullet$\end{tabular}}}}%
    \put(0.90990272,0.25922712){\makebox(0,0)[lt]{\lineheight{1.25}\smash{\begin{tabular}[t]{l}$M$\end{tabular}}}}%
  \end{picture}%
\endgroup%

	\caption{Assignment in ASG OS}
	\label{fig:asgr}
\end{wrapfigure}
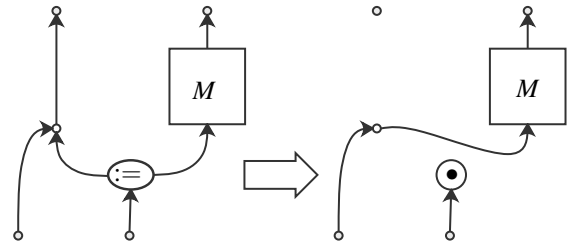

Fig~\ref{fig:asgr} shows the basic rule for assignment, with the atom indicated as an unlabeled white node. 
The atom is made to point to the second operand of the assignment operator, while the assignment operator itself reduces to the dummy value inhabiting the unit type. 
In the process, whatever the atom was attached to before may become inaccessible from the root of the ASG, therefore \textit{garbage.}
Also note that the effect of the assignment is manifest only because other parts of the ASG may point to the atom, a link which is persistent due to the value-like behaviour of the atom. 

The SOS of a programming language can be further refined (\textit{distilled}) into an abstract machine, which gives a more explicit representation of the simplification rules via manipulation of context~\cite{DBLP:journals/iandc/WrightF94}.
From this point of view, the ASG representation of the SOS is already an abstract machine, in the sense that it can give a cost-accurate model of execution of the language. 

Another appealing feature of abstract machines is that they can model control-transfer operations more conveniently that SOS.
It is not impossible to use SOS for this, but the format of the transition system needs to be significantly revised, making the transitions themselves labelled~\cite{DBLP:journals/corr/SculthorpeTM16}.
\begin{figure}
	\centering\small 
	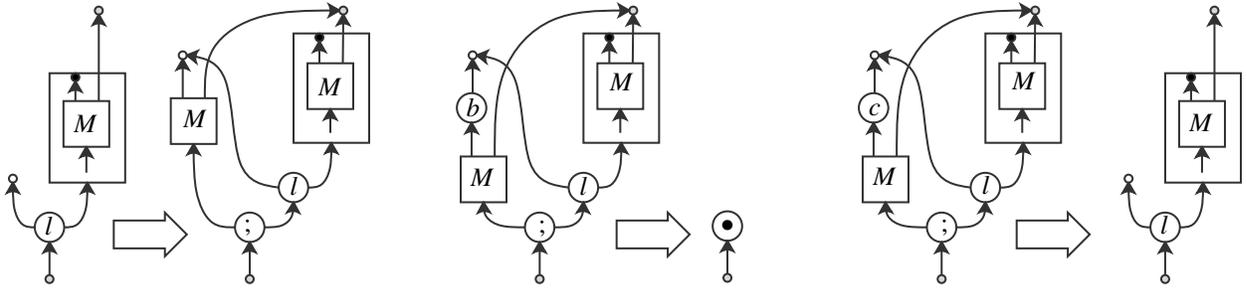
	\caption{Control in ASG OS}
	\label{fig:control}
\end{figure}

Since the ASG OS is formulated via arbitrary rewrites, control can be dealt with in a straightforward way. 
Fig~\ref{fig:control} shows a labelled version of C-style \textit{break/continue} statements.
The operations involved are loop body definition ($l$), sequential composition (;), break ($b$), and continue ($c$). 
The atom used as the first operand of $l$ becomes bound to the label which is the bound argument of $M$, used to anchor a point in the ASG so that the control operations of break or continue can determine where to jump to. 
If $M$ terminates normally then the whole cycle repeats.
Unlike conventional C break and continue these variants are labelled, and the labels are first-class citizens, i.e. they can be passed as arguments to or returned from functions. 

An interactive evaluator for a variety of programming language features can be found online.\footnote{\url{https://tnttodda.github.io/Spartan-Visualiser/}}.

\section{Reasoning}

SOS was originally considered too `low level' to reason about equivalence in programming languages, at least in contrast with denotational semantics. 
However, SOS was considered more `high level' than alternative operational specifications of programming languages such as abstract machines. 
In time, a large variety of powerful techniques for reasoning with SOS-like specifications proved that this is indeed a useful formalism for reasoning about equivalence~\cite{DBLP:conf/ac/Pitts00} whereas abstract machines remained useful due to their ability to model the cost of evaluation and as a gateway to compiler development~\cite{leroy:inria-00070049}. 
The ASG OS seems to combine felicitously some of the best features of SOS and abstract machines including, as we shall see, the ability to reason about observational equivalence. 

In fact, the graph formulation of the OS makes it possible not just to reason about equivalence, but to formulate a powerful characterisation theorem which establishes equivalence by using some simpler combinatorial criteria. 
We must first `tame' the OS by restricting ourselves to sets of rules which are \textit{deterministic} and \textit{refocussing}. 
The first concept is the standard one. 
The second, initially formulated by Danvy et. al., means that following a basic reduction the focus could be kept either at the point where the rewrite occurs, or moved to the root of the graph, with equal effect~\cite{DBLP:journals/tcs/DanvyMMZ12}.
Indeed, all the rules we have presented in this tutorial are refocussing.  

Equivalences are also formulated graphically, as families of relations on \textit{templates}, i.e. sets of graphs with the same interface. 
For a fixed abstract machine a template is said to be \textit{input-safe} if evaluation from any input link preserves the relation. 
Note that, unlike a SOS, we can talk about the evaluation of an ASG which is not a program, in fact not even a term, since evaluation is just a byword for traversal and reduction. 
A template is said to be \textit{output-closed} if in the course of evaluation no output link will ever be reached. 
Finally, a template is said to be \textit{robust} if it is preserved by all rewrite rules of the language. 
The main theorem can be simply stated as:
\\[2ex]
\textbf{Theorem.} (Characterisation~\cite[Sec.~6]{DBLP:journals/corr/abs-1907-01257}). \textit{Robust templates induce observational equivalence. }
\\[2ex]
The conditions used to establish equivalence via the Characterisation Theorem are all elementary and proved by case analysis. 
Moreover, the theorem allows for \textit{robust} proofs of equivalence in the sense that they can withstand language extensions. 
For example we can prove the $\beta$ law for a pure language can be extended to a language with imperative store just by showing that the templates used in formulating the law are robust relative to the new rules for variable creation, dereferencing, and assignment (Fig.~\ref{fig:asg}).  
Which happens to be the case. 
By contrast, conventional proofs of equivalence are fragile, and are invalidated by even mild language extensions. 

\section{Related work}

This is an elementary tutorial introduction and extended motivation for the \textit{hypernet semantics} of programming languages~\cite{DBLP:journals/corr/abs-1907-01257}, which is a streamlined and generalised version of the \textit{Dynamic Geometry of Interaction (GoI) Machine}~\cite{DBLP:journals/lmcs/MuroyaG19}.
They are the outcome of an effort initially motivated by the understanding of  call-by-value and effectful computation from a GoI perspective~\cite{DBLP:conf/csl/HoshinoMH14,DBLP:conf/popl/MuroyaHH16}.

Graph-based intermediate representations (IR) are established in compiler construction~\cite{DBLP:conf/irep/ClickP95} and in the formulation of abstract machines for functional languages~\cite{DBLP:conf/fpca/JonesS89}.
However, the origin of the approach describe here lies elsewhere, in \textit{proof nets}, a graphical representation of proofs in Linear Logic~\cite{DBLP:journals/tcs/Girard87} and especially in their generalisation as \textit{interaction nets}~\cite{DBLP:conf/popl/Lafont90}.
Interaction nets already exhibit the hierarchical structure we employ here, which is used to model binding and higher-order structures. 
Hierarchical graphs are also used elsewhere in semantics, for example as diagram languages of processes known as \textit{bigraphs}~\cite{DBLP:journals/entcs/Milner08}. 

The connection between linear logic and its varieties and certain monoidal categories kindled significant progress in diagrammatic languages~\cite{Selinger2011}.
For instance, traced monoidal categories, used as models of lambda calculus with cyclic sharing~\cite{DBLP:conf/tlca/Hasegawa97}, led to the development of a hierarchical graph syntax for closures~\cite{DBLP:journals/entcs/SchweimeierJ99} remarkably similar to the one described here.
In terms of the treatment of graphs as combinatorial objects, much of the literature considered them rather informally and a formalisation of proof nets as hypergraphs was given much later~\cite{DBLP:journals/tcs/GuerriniMM01}.

More recently, work by Accattoli has examined the interesting interplay between term-based and graph-based formulations of the call-by-value lambda calculus~\cite{DBLP:journals/tcs/Accattoli15}, even though his motivations are somewhat different than ours, as illustrated by this quotation:
\begin{quotation}\em
	It is far from easy to realize an isomorphism
	between terms and nets, as it is necessary
	to take care of many delicate details about
	weakenings, contractions, representation of
	variables, administrative reduction steps,
	and so on. [$\ldots$] More generally, such a strong
	relationship turns the calculus into an
	algebraic language for proof nets, providing
	a handy tool to reason by structural
	induction over proof nets.
\end{quotation}
In fact, a properly formalised graphical syntax can be just as powerful and just as rigorous as an algebraic language. 
Moreover, the graphical language can be both simpler and better specified than the term language, for example in the case of the calculus of explicit substitutions, which lacks a proper formulation of $\alpha$-equivalence~\cite{DBLP:conf/ictac/Accattoli18}.

To conclude, we see the ASG operational semantics as a first step in an exciting and potentially fruitful direction. 
Graphical languages are starting to emerge as a new and genuine formalism which can give alternative, and sometimes improved, representations to theories in fields as different as quantum computation~\cite{coecke_kissinger_2017}, linear and affine algebra~\cite{DBLP:conf/lics/BonchiPSZ19}, digital circuits~\cite{DBLP:conf/csl/GhicaJL17}, signal flow~\cite{DBLP:conf/popl/BonchiSZ15} and more. 
The motivations for this emergence are mixed, from the raw intuitive appeal of visual representations to improved algorithmic properties. 
Examining how this methodology can be extended to programming languages is an intriguing next step which brings together a number of existing ideas and concepts and can unify existing gaps between semantics of programming languages and compiler-related techniques. 

\bibliographystyle{eptcs}
\bibliography{generic}

\begin{thebibliography}{10}
\providecommand{\bibitemdeclare}[2]{}
\providecommand{\surnamestart}{}
\providecommand{\surnameend}{}
\providecommand{\urlprefix}{Available at }
\providecommand{\url}[1]{\texttt{#1}}
\providecommand{\href}[2]{\texttt{#2}}
\providecommand{\urlalt}[2]{\href{#1}{#2}}
\providecommand{\doi}[1]{doi:\urlalt{http://dx.doi.org/#1}{#1}}
\providecommand{\bibinfo}[2]{#2}

\bibitemdeclare{article}{DBLP:journals/jfp/AbadiCCL91}
\bibitem{DBLP:journals/jfp/AbadiCCL91}
\bibinfo{author}{Mart{\'{\i}}n \surnamestart Abadi\surnameend},
  \bibinfo{author}{Luca \surnamestart Cardelli\surnameend},
  \bibinfo{author}{Pierre{-}Louis \surnamestart Curien\surnameend} \&
  \bibinfo{author}{Jean{-}Jacques \surnamestart L{\'{e}}vy\surnameend}
  (\bibinfo{year}{1991}): \emph{\bibinfo{title}{Explicit Substitutions}}.
\newblock {\sl \bibinfo{journal}{J. Funct. Program.}}
  \bibinfo{volume}{1}(\bibinfo{number}{4}), pp. \bibinfo{pages}{375--416},
  \doi{10.1017/S0956796800000186}.

\bibitemdeclare{article}{DBLP:journals/tcs/Accattoli15}
\bibitem{DBLP:journals/tcs/Accattoli15}
\bibinfo{author}{Beniamino \surnamestart Accattoli\surnameend}
  (\bibinfo{year}{2015}): \emph{\bibinfo{title}{Proof nets and the
  call-by-value {\(\lambda\)}-calculus}}.
\newblock {\sl \bibinfo{journal}{Theor. Comput. Sci.}} \bibinfo{volume}{606},
  pp. \bibinfo{pages}{2--24}, \doi{10.1016/j.tcs.2015.08.006}.

\bibitemdeclare{inproceedings}{DBLP:conf/ictac/Accattoli18}
\bibitem{DBLP:conf/ictac/Accattoli18}
\bibinfo{author}{Beniamino \surnamestart Accattoli\surnameend}
  (\bibinfo{year}{2018}): \emph{\bibinfo{title}{Proof Nets and the Linear
  Substitution Calculus}}.
\newblock In \bibinfo{editor}{Bernd \surnamestart Fischer\surnameend} \&
  \bibinfo{editor}{Tarmo \surnamestart Uustalu\surnameend}, editors: {\sl
  \bibinfo{booktitle}{Theoretical Aspects of Computing - {ICTAC} 2018 - 15th
  International Colloquium, Stellenbosch, South Africa, October 16-19, 2018,
  Proceedings}}, {\sl \bibinfo{series}{Lecture Notes in Computer Science}}
  \bibinfo{volume}{11187}, \bibinfo{publisher}{Springer}, pp.
  \bibinfo{pages}{37--61}, \doi{10.1007/978-3-030-02508-3\_3}.

\bibitemdeclare{inproceedings}{DBLP:conf/popl/AccattoliBKL14}
\bibitem{DBLP:conf/popl/AccattoliBKL14}
\bibinfo{author}{Beniamino \surnamestart Accattoli\surnameend},
  \bibinfo{author}{Eduardo \surnamestart Bonelli\surnameend},
  \bibinfo{author}{Delia \surnamestart Kesner\surnameend} \&
  \bibinfo{author}{Carlos \surnamestart Lombardi\surnameend}
  (\bibinfo{year}{2014}): \emph{\bibinfo{title}{A nonstandard standardization
  theorem}}.
\newblock In \bibinfo{editor}{Suresh \surnamestart Jagannathan\surnameend} \&
  \bibinfo{editor}{Peter \surnamestart Sewell\surnameend}, editors: {\sl
  \bibinfo{booktitle}{The 41st Annual {ACM} {SIGPLAN-SIGACT} Symposium on
  Principles of Programming Languages, {POPL} '14, San Diego, CA, USA, January
  20-21, 2014}}, \bibinfo{publisher}{{ACM}}, pp. \bibinfo{pages}{659--670},
  \doi{10.1145/2535838.2535886}.

\bibitemdeclare{inproceedings}{DBLP:conf/lics/BonchiPSZ19}
\bibitem{DBLP:conf/lics/BonchiPSZ19}
\bibinfo{author}{Filippo \surnamestart Bonchi\surnameend},
  \bibinfo{author}{Robin \surnamestart Piedeleu\surnameend},
  \bibinfo{author}{Pawel \surnamestart Sobocinski\surnameend} \&
  \bibinfo{author}{Fabio \surnamestart Zanasi\surnameend}
  (\bibinfo{year}{2019}): \emph{\bibinfo{title}{Graphical Affine Algebra}}.
\newblock In: {\sl \bibinfo{booktitle}{34th Annual {ACM/IEEE} Symposium on
  Logic in Computer Science, {LICS} 2019, Vancouver, BC, Canada, June 24-27,
  2019}}, \bibinfo{publisher}{{IEEE}}, pp. \bibinfo{pages}{1--12},
  \doi{10.1109/LICS.2019.8785877}.

\bibitemdeclare{inproceedings}{DBLP:conf/popl/BonchiSZ15}
\bibitem{DBLP:conf/popl/BonchiSZ15}
\bibinfo{author}{Filippo \surnamestart Bonchi\surnameend},
  \bibinfo{author}{Pawel \surnamestart Sobocinski\surnameend} \&
  \bibinfo{author}{Fabio \surnamestart Zanasi\surnameend}
  (\bibinfo{year}{2015}): \emph{\bibinfo{title}{Full Abstraction for Signal
  Flow Graphs}}.
\newblock In \bibinfo{editor}{Sriram~K. \surnamestart Rajamani\surnameend} \&
  \bibinfo{editor}{David \surnamestart Walker\surnameend}, editors: {\sl
  \bibinfo{booktitle}{Proceedings of the 42nd Annual {ACM} {SIGPLAN-SIGACT}
  Symposium on Principles of Programming Languages, {POPL} 2015, Mumbai, India,
  January 15-17, 2015}}, \bibinfo{publisher}{{ACM}}, pp.
  \bibinfo{pages}{515--526}, \doi{10.1145/2676726.2676993}.

\bibitemdeclare{inproceedings}{DBLP:conf/irep/ClickP95}
\bibitem{DBLP:conf/irep/ClickP95}
\bibinfo{author}{Cliff \surnamestart Click\surnameend} \&
  \bibinfo{author}{Michael \surnamestart Paleczny\surnameend}
  (\bibinfo{year}{1995}): \emph{\bibinfo{title}{A Simple Graph-Based
  Intermediate Representation}}.
\newblock In \bibinfo{editor}{Michael~D. \surnamestart Ernst\surnameend},
  editor: {\sl \bibinfo{booktitle}{Proceedings {ACM} {SIGPLAN} Workshop on
  Intermediate Representations (IR'95), San Francisco, CA, USA, January 22,
  1995}}, \bibinfo{publisher}{{ACM}}, pp. \bibinfo{pages}{35--49},
  \doi{10.1145/202529.202534}.

\bibitemdeclare{book}{coecke_kissinger_2017}
\bibitem{coecke_kissinger_2017}
\bibinfo{author}{Bob \surnamestart Coecke\surnameend} \& \bibinfo{author}{Aleks
  \surnamestart Kissinger\surnameend} (\bibinfo{year}{2017}):
  \emph{\bibinfo{title}{Picturing Quantum Processes: A First Course in Quantum
  Theory and Diagrammatic Reasoning}}.
\newblock \bibinfo{publisher}{Cambridge University Press},
  \doi{10.1017/9781316219317}.

\bibitemdeclare{article}{DBLP:journals/tcs/DanvyMMZ12}
\bibitem{DBLP:journals/tcs/DanvyMMZ12}
\bibinfo{author}{Olivier \surnamestart Danvy\surnameend},
  \bibinfo{author}{Kevin \surnamestart Millikin\surnameend},
  \bibinfo{author}{Johan \surnamestart Munk\surnameend} \& \bibinfo{author}{Ian
  \surnamestart Zerny\surnameend} (\bibinfo{year}{2012}):
  \emph{\bibinfo{title}{On inter-deriving small-step and big-step semantics:
  {A} case study for storeless call-by-need evaluation}}.
\newblock {\sl \bibinfo{journal}{Theor. Comput. Sci.}} \bibinfo{volume}{435},
  pp. \bibinfo{pages}{21--42}, \doi{10.1016/j.tcs.2012.02.023}.

\bibitemdeclare{inproceedings}{DBLP:conf/csl/GhicaJL17}
\bibitem{DBLP:conf/csl/GhicaJL17}
\bibinfo{author}{Dan~R. \surnamestart Ghica\surnameend}, \bibinfo{author}{Achim
  \surnamestart Jung\surnameend} \& \bibinfo{author}{Aliaume \surnamestart
  Lopez\surnameend} (\bibinfo{year}{2017}): \emph{\bibinfo{title}{Diagrammatic
  Semantics for Digital Circuits}}.
\newblock In \bibinfo{editor}{Valentin \surnamestart Goranko\surnameend} \&
  \bibinfo{editor}{Mads \surnamestart Dam\surnameend}, editors: {\sl
  \bibinfo{booktitle}{26th {EACSL} Annual Conference on Computer Science Logic,
  {CSL} 2017, August 20-24, 2017, Stockholm, Sweden}}, {\sl
  \bibinfo{series}{LIPIcs}}~\bibinfo{volume}{82}, \bibinfo{publisher}{Schloss
  Dagstuhl - Leibniz-Zentrum f{\"{u}}r Informatik}, pp.
  \bibinfo{pages}{24:1--24:16}, \doi{10.4230/LIPIcs.CSL.2017.24}.

\bibitemdeclare{article}{DBLP:journals/corr/abs-1907-01257}
\bibitem{DBLP:journals/corr/abs-1907-01257}
\bibinfo{author}{Dan~R. \surnamestart Ghica\surnameend}, \bibinfo{author}{Koko
  \surnamestart Muroya\surnameend} \& \bibinfo{author}{Todd~Waugh \surnamestart
  Ambridge\surnameend} (\bibinfo{year}{2019}): \emph{\bibinfo{title}{Local
  Reasoning for Robust Observational Equivalence}}.
\newblock {\sl \bibinfo{journal}{CoRR}} \bibinfo{volume}{abs/1907.01257}.
\newblock \urlprefix\url{http://arxiv.org/abs/1907.01257}.

\bibitemdeclare{inproceedings}{DBLP:journals/entcs/GhicaT12}
\bibitem{DBLP:journals/entcs/GhicaT12}
\bibinfo{author}{Dan~R. \surnamestart Ghica\surnameend} \&
  \bibinfo{author}{Nikos \surnamestart Tzevelekos\surnameend}
  (\bibinfo{year}{2012}): \emph{\bibinfo{title}{A System-Level Game
  Semantics}}.
\newblock In \bibinfo{editor}{Ulrich \surnamestart Berger\surnameend} \&
  \bibinfo{editor}{Michael~W. \surnamestart Mislove\surnameend}, editors: {\sl
  \bibinfo{booktitle}{Proceedings of the 28th Conference on the Mathematical
  Foundations of Programming Semantics, {MFPS} 2012, Bath, UK, June 6-9,
  2012}}, {\sl \bibinfo{series}{Electronic Notes in Theoretical Computer
  Science}} \bibinfo{volume}{286}, \bibinfo{publisher}{Elsevier}, pp.
  \bibinfo{pages}{191--211}, \doi{10.1016/j.entcs.2012.08.013}.

\bibitemdeclare{article}{DBLP:journals/tcs/Girard87}
\bibitem{DBLP:journals/tcs/Girard87}
\bibinfo{author}{Jean{-}Yves \surnamestart Girard\surnameend}
  (\bibinfo{year}{1987}): \emph{\bibinfo{title}{Linear Logic}}.
\newblock {\sl \bibinfo{journal}{Theor. Comput. Sci.}} \bibinfo{volume}{50},
  pp. \bibinfo{pages}{1--102}, \doi{10.1016/0304-3975(87)90045-4}.

\bibitemdeclare{article}{DBLP:journals/tcs/GuerriniMM01}
\bibitem{DBLP:journals/tcs/GuerriniMM01}
\bibinfo{author}{Stefano \surnamestart Guerrini\surnameend},
  \bibinfo{author}{Simone \surnamestart Martini\surnameend} \&
  \bibinfo{author}{Andrea \surnamestart Masini\surnameend}
  (\bibinfo{year}{2001}): \emph{\bibinfo{title}{Proof nets, garbage, and
  computations}}.
\newblock {\sl \bibinfo{journal}{Theor. Comput. Sci.}}
  \bibinfo{volume}{253}(\bibinfo{number}{2}), pp. \bibinfo{pages}{185--237},
  \doi{10.1016/S0304-3975(00)00094-3}.

\bibitemdeclare{inproceedings}{DBLP:conf/tlca/Hasegawa97}
\bibitem{DBLP:conf/tlca/Hasegawa97}
\bibinfo{author}{Masahito \surnamestart Hasegawa\surnameend}
  (\bibinfo{year}{1997}): \emph{\bibinfo{title}{Recursion from Cyclic Sharing:
  Traced Monoidal Categories and Models of Cyclic Lambda Calculi}}.
\newblock In \bibinfo{editor}{Philippe \surnamestart de~Groote\surnameend},
  editor: {\sl \bibinfo{booktitle}{Typed Lambda Calculi and Applications, Third
  International Conference on Typed Lambda Calculi and Applications, {TLCA}
  '97, Nancy, France, April 2-4, 1997, Proceedings}}, {\sl
  \bibinfo{series}{Lecture Notes in Computer Science}} \bibinfo{volume}{1210},
  \bibinfo{publisher}{Springer}, pp. \bibinfo{pages}{196--213},
  \doi{10.1007/3-540-62688-3\_37}.

\bibitemdeclare{inproceedings}{DBLP:conf/csl/HoshinoMH14}
\bibitem{DBLP:conf/csl/HoshinoMH14}
\bibinfo{author}{Naohiko \surnamestart Hoshino\surnameend},
  \bibinfo{author}{Koko \surnamestart Muroya\surnameend} \&
  \bibinfo{author}{Ichiro \surnamestart Hasuo\surnameend}
  (\bibinfo{year}{2014}): \emph{\bibinfo{title}{Memoryful geometry of
  interaction: from coalgebraic components to algebraic effects}}.
\newblock In \bibinfo{editor}{Thomas~A. \surnamestart Henzinger\surnameend} \&
  \bibinfo{editor}{Dale \surnamestart Miller\surnameend}, editors: {\sl
  \bibinfo{booktitle}{Joint Meeting of the Twenty-Third {EACSL} Annual
  Conference on Computer Science Logic {(CSL)} and the Twenty-Ninth Annual
  {ACM/IEEE} Symposium on Logic in Computer Science (LICS), {CSL-LICS} '14,
  Vienna, Austria, July 14 - 18, 2014}}, \bibinfo{publisher}{{ACM}}, pp.
  \bibinfo{pages}{52:1--52:10}, \doi{10.1145/2603088.2603124}.

\bibitemdeclare{inproceedings}{DBLP:conf/fpca/JonesS89}
\bibitem{DBLP:conf/fpca/JonesS89}
\bibinfo{author}{Simon L.~Peyton \surnamestart Jones\surnameend} \&
  \bibinfo{author}{Jon \surnamestart Salkild\surnameend}
  (\bibinfo{year}{1989}): \emph{\bibinfo{title}{The Spineless Tagless
  G-Machine}}.
\newblock In \bibinfo{editor}{Joseph~E. \surnamestart Stoy\surnameend}, editor:
  {\sl \bibinfo{booktitle}{Proceedings of the fourth international conference
  on Functional programming languages and computer architecture, {FPCA} 1989,
  London, UK, September 11-13, 1989}}, \bibinfo{publisher}{{ACM}}, pp.
  \bibinfo{pages}{184--201}, \doi{10.1145/99370.99385}.

\bibitemdeclare{inproceedings}{DBLP:conf/popl/Lafont90}
\bibitem{DBLP:conf/popl/Lafont90}
\bibinfo{author}{Yves \surnamestart Lafont\surnameend} (\bibinfo{year}{1990}):
  \emph{\bibinfo{title}{Interaction Nets}}.
\newblock In \bibinfo{editor}{Frances~E. \surnamestart Allen\surnameend},
  editor: {\sl \bibinfo{booktitle}{Conference Record of the Seventeenth Annual
  {ACM} Symposium on Principles of Programming Languages, San Francisco,
  California, USA, January 1990}}, \bibinfo{publisher}{{ACM} Press}, pp.
  \bibinfo{pages}{95--108}, \doi{10.1145/96709.96718}.

\bibitemdeclare{techreport}{leroy:inria-00070049}
\bibitem{leroy:inria-00070049}
\bibinfo{author}{Xavier \surnamestart Leroy\surnameend} (\bibinfo{year}{1990}):
  \emph{\bibinfo{title}{{The ZINC experiment : an economical implementation of
  the ML language}}}.
\newblock \bibinfo{type}{Technical Report} \bibinfo{number}{RT-0117},
  \bibinfo{institution}{{INRIA}}.
\newblock \urlprefix\url{https://hal.inria.fr/inria-00070049}.

\bibitemdeclare{article}{DBLP:journals/entcs/Milner08}
\bibitem{DBLP:journals/entcs/Milner08}
\bibinfo{author}{Robin \surnamestart Milner\surnameend} (\bibinfo{year}{2008}):
  \emph{\bibinfo{title}{Bigraphs and Their Algebra}}.
\newblock {\sl \bibinfo{journal}{Electron. Notes Theor. Comput. Sci.}}
  \bibinfo{volume}{209}, pp. \bibinfo{pages}{5--19},
  \doi{10.1016/j.entcs.2008.04.002}.

\bibitemdeclare{phdthesis}{koko}
\bibitem{koko}
\bibinfo{author}{Koko \surnamestart Muroya\surnameend} (\bibinfo{year}{2020}):
  \emph{\bibinfo{title}{Hypernet Semantics of Programming Languages}}.
\newblock Ph.D. thesis, \bibinfo{school}{University of Birmingham}.
\newblock
  \urlprefix\url{http://www.kurims.kyoto-u.ac.jp/~kmuroya/papers/phdthesis.pdf}.

\bibitemdeclare{article}{DBLP:journals/lmcs/MuroyaG19}
\bibitem{DBLP:journals/lmcs/MuroyaG19}
\bibinfo{author}{Koko \surnamestart Muroya\surnameend} \&
  \bibinfo{author}{Dan~R. \surnamestart Ghica\surnameend}
  (\bibinfo{year}{2019}): \emph{\bibinfo{title}{The Dynamic Geometry of
  Interaction Machine: {A} Token-Guided Graph Rewriter}}.
\newblock {\sl \bibinfo{journal}{Log. Methods Comput. Sci.}}
  \bibinfo{volume}{15}(\bibinfo{number}{4}).
\newblock \urlprefix\url{https://lmcs.episciences.org/5882}.

\bibitemdeclare{inproceedings}{DBLP:conf/popl/MuroyaHH16}
\bibitem{DBLP:conf/popl/MuroyaHH16}
\bibinfo{author}{Koko \surnamestart Muroya\surnameend},
  \bibinfo{author}{Naohiko \surnamestart Hoshino\surnameend} \&
  \bibinfo{author}{Ichiro \surnamestart Hasuo\surnameend}
  (\bibinfo{year}{2016}): \emph{\bibinfo{title}{Memoryful geometry of
  interaction {II:} recursion and adequacy}}.
\newblock In \bibinfo{editor}{Rastislav \surnamestart Bod{\'{\i}}k\surnameend}
  \& \bibinfo{editor}{Rupak \surnamestart Majumdar\surnameend}, editors: {\sl
  \bibinfo{booktitle}{Proceedings of the 43rd Annual {ACM} {SIGPLAN-SIGACT}
  Symposium on Principles of Programming Languages, {POPL} 2016, St.
  Petersburg, FL, USA, January 20 - 22, 2016}}, \bibinfo{publisher}{{ACM}}, pp.
  \bibinfo{pages}{748--760}, \doi{10.1145/2837614.2837672}.

\bibitemdeclare{inproceedings}{DBLP:conf/lics/Pitts96}
\bibitem{DBLP:conf/lics/Pitts96}
\bibinfo{author}{Andrew~M. \surnamestart Pitts\surnameend}
  (\bibinfo{year}{1996}): \emph{\bibinfo{title}{Reasoning about Local Variables
  with Operationally-Based Logical Relations}}.
\newblock In: {\sl \bibinfo{booktitle}{Proceedings, 11th Annual {IEEE}
  Symposium on Logic in Computer Science, New Brunswick, New Jersey, USA, July
  27-30, 1996}}, \bibinfo{publisher}{{IEEE} Computer Society}, pp.
  \bibinfo{pages}{152--163}, \doi{10.1109/LICS.1996.561314}.

\bibitemdeclare{inproceedings}{DBLP:conf/ac/Pitts00}
\bibitem{DBLP:conf/ac/Pitts00}
\bibinfo{author}{Andrew~M. \surnamestart Pitts\surnameend}
  (\bibinfo{year}{2000}): \emph{\bibinfo{title}{Operational Semantics and
  Program Equivalence}}.
\newblock In \bibinfo{editor}{Gilles \surnamestart Barthe\surnameend},
  \bibinfo{editor}{Peter \surnamestart Dybjer\surnameend},
  \bibinfo{editor}{Lu{\'{\i}}s \surnamestart Pinto\surnameend} \&
  \bibinfo{editor}{Jo{\~{a}}o \surnamestart Saraiva\surnameend}, editors: {\sl
  \bibinfo{booktitle}{Applied Semantics, International Summer School, {APPSEM}
  2000, Caminha, Portugal, September 9-15, 2000, Advanced Lectures}}, {\sl
  \bibinfo{series}{Lecture Notes in Computer Science}} \bibinfo{volume}{2395},
  \bibinfo{publisher}{Springer}, pp. \bibinfo{pages}{378--412},
  \doi{10.1007/3-540-45699-6\_8}.

\bibitemdeclare{book}{pitts_2013}
\bibitem{pitts_2013}
\bibinfo{author}{Andrew~M. \surnamestart Pitts\surnameend}
  (\bibinfo{year}{2013}): \emph{\bibinfo{title}{Nominal Sets: Names and
  Symmetry in Computer Science}}.
\newblock \bibinfo{series}{Cambridge Tracts in Theoretical Computer Science},
  \bibinfo{publisher}{Cambridge University Press},
  \doi{10.1017/CBO9781139084673}.

\bibitemdeclare{article}{DBLP:journals/jlp/Plotkin04a}
\bibitem{DBLP:journals/jlp/Plotkin04a}
\bibinfo{author}{Gordon~D. \surnamestart Plotkin\surnameend}
  (\bibinfo{year}{2004}): \emph{\bibinfo{title}{A structural approach to
  operational semantics}}.
\newblock {\sl \bibinfo{journal}{J. Log. Algebraic Methods Program.}}
  \bibinfo{volume}{60-61}, pp. \bibinfo{pages}{17--139}
  \doi{10.1016/j.jlap.2004.05.001}.

\bibitemdeclare{inproceedings}{DBLP:journals/entcs/SchweimeierJ99}
\bibitem{DBLP:journals/entcs/SchweimeierJ99}
\bibinfo{author}{Ralf \surnamestart Schweimeier\surnameend} \&
  \bibinfo{author}{Alan \surnamestart Jeffrey\surnameend}
  (\bibinfo{year}{1999}): \emph{\bibinfo{title}{A Categorical and Graphical
  Treatment of Closure Conversion}}.
\newblock In \bibinfo{editor}{Stephen~D. \surnamestart Brookes\surnameend},
  \bibinfo{editor}{Achim \surnamestart Jung\surnameend},
  \bibinfo{editor}{Michael~W. \surnamestart Mislove\surnameend} \&
  \bibinfo{editor}{Andre \surnamestart Scedrov\surnameend}, editors: {\sl
  \bibinfo{booktitle}{Fifteenth Conference on Mathematical Foundations of
  Progamming Semantics, {MFPS} 1999, Tulane University, New Orleans, LA, USA,
  April 28 - May 1, 1999}}, {\sl \bibinfo{series}{Electronic Notes in
  Theoretical Computer Science}}~\bibinfo{volume}{20},
  \bibinfo{publisher}{Elsevier}, pp. \bibinfo{pages}{481--511},
  \doi{10.1016/S1571-0661(04)80090-2}.

\bibitemdeclare{inproceedings}{DBLP:journals/corr/SculthorpeTM16}
\bibitem{DBLP:journals/corr/SculthorpeTM16}
\bibinfo{author}{Neil \surnamestart Sculthorpe\surnameend},
  \bibinfo{author}{Paolo \surnamestart Torrini\surnameend} \&
  \bibinfo{author}{Peter~D. \surnamestart Mosses\surnameend}
  (\bibinfo{year}{2015}): \emph{\bibinfo{title}{A Modular Structural
  Operational Semantics for Delimited Continuations}}.
\newblock In \bibinfo{editor}{Olivier \surnamestart Danvy\surnameend} \&
  \bibinfo{editor}{Ugo \surnamestart de'Liguoro\surnameend}, editors: {\sl
  \bibinfo{booktitle}{Proceedings of the Workshop on Continuations, WoC 2016,
  London, UK, April 12th 2015}}, {\sl \bibinfo{series}{{EPTCS}}}
  \bibinfo{volume}{212}, pp. \bibinfo{pages}{63--80},
  \doi{10.4204/EPTCS.212.5}.

\bibitemdeclare{inbook}{Selinger2011}
\bibitem{Selinger2011}
\bibinfo{author}{P.~\surnamestart Selinger\surnameend} (\bibinfo{year}{2011}):
  \emph{\bibinfo{title}{A Survey of Graphical Languages for Monoidal
  Categories}}, pp. \bibinfo{pages}{289--355}.
\newblock \bibinfo{publisher}{Springer Berlin Heidelberg},
  \bibinfo{address}{Berlin, Heidelberg}, \doi{10.1007/978-3-642-12821-9\_4}.

\bibitemdeclare{article}{DBLP:journals/iandc/WrightF94}
\bibitem{DBLP:journals/iandc/WrightF94}
\bibinfo{author}{Andrew~K. \surnamestart Wright\surnameend} \&
  \bibinfo{author}{Matthias \surnamestart Felleisen\surnameend}
  (\bibinfo{year}{1994}): \emph{\bibinfo{title}{A Syntactic Approach to Type
  Soundness}}.
\newblock {\sl \bibinfo{journal}{Inf. Comput.}}
  \bibinfo{volume}{115}(\bibinfo{number}{1}), pp. \bibinfo{pages}{38--94},
  \doi{10.1006/inco.1994.1093}.

\end{thebibliography}
\end{document}